\documentclass{elsart}

\usepackage{natbib,epsfig}

\begin{document}

\runauthor{name of author(s)}


\begin{frontmatter}


\title{  CP and T violation in (long) 
long baseline neutrino oscillation
experiments
}


\author{
Joe Sato
}
\address {
 Department of Physics, University of Tokyo, Hongo, Bunkyo-ku, Tokyo,
   113-0033, Japan%
}
\thanks[currentaddress]{Current address: Research Center for Higher
Education, Kyushu University,\\ Ropponmatsu, Chuo-ku, Fukuoka, 810-8560,
Japan}






\begin{abstract}
We consider possibilities of observing
CP-violation effects in neutrino oscillation experiments
with low energy ($\sim$ several hundreds MeV).
\end{abstract}




\end{frontmatter}






\section{Introduction}

Many experiments and observations have shown evidences for neutrino
oscillation one after another.  The solar neutrino deficit has long
been observed\cite{Ga1,Ga2,Kam,Cl,SolSK}.  The atmospheric neutrino
anomaly has been found\cite{AtmKam,IMB,SOUDAN2,MACRO} and recently
almost confirmed by SuperKamiokande\cite{AtmSK}.  There is also
another suggestion given by LSND\cite{LSND}.  All of them can be
understood by neutrino oscillation and hence indicates that neutrinos
are massive and there is a mixing in lepton
sector\cite{FukugitaYanagida}.

Since there is a mixing in lepton sector, it is quite natural to
imagine that there occurs CP violation in lepton sector.  Several
physicists have considered whether we may see CP-violation effect in
lepton sector through long baseline neutrino oscillation experiments. 
First it has been studied in the context of currently planed
experiments\cite{Tanimoto,ArafuneJoe,AKS,MN,BGG} and recently in the
context of neutrino factory\cite{BGW,Tanimoto2,Romanino,GH}.

The use of neutrinos from muon beam has great advantages compared with
those from pion beam\cite{Geer}.
Neutrinos from $\mu^+$($\mu^-$) beam consist of
pure $\nu_{\rm e}$ and $\bar\nu_\mu$ ($\bar\nu_{\rm e}$ and $\nu_\mu$)
and will contain no contamination of other kinds of neutrinos.  Also
their energy distribution will be determined very well.

In this proceedings,
we will consider how large CP-violation effect we will see
in oscillation experiments with low energy neutrino
from muon beam. Such neutrinos with high intensity will be available
in near future\cite{PRISM}.
We will consider three active neutrinos without any
sterile one by attributing the solar neutrino deficit and atmospheric
neutrino anomaly to the neutrino oscillation.

\section{CP violation in long baseline neutrino oscillation experiments}

Here we consider neutrino oscillation experiments with baseline
$L\sim$ several hundreds km.

\subsection{Oscillation probability and its approximated formula}

First we derive approximated formulas\cite{AKS} of neutrino oscillation
to clarify our notation.

We assume three generations of neutrinos which have mass eigenvalues
$m_{i} (i=1, 2, 3)$ and MNS mixing matrix $U$\cite{MNS} relating the flavor
eigenstates $\nu_{\alpha} (\alpha={\rm e}, \mu, \tau)$ and the mass
eigenstates in the vacuum $\nu\,'_{i} (i=1, 2, 3)$ as
\begin{equation}
  \nu_{\alpha} = U_{\alpha i} \nu\,'_{i}.
  \label{Udef}
\end{equation}
We parameterize $U$\cite{ChauKeung,KuoPnataleone,Toshev}
as

\begin{eqnarray}
& &
U
=
{\rm e}^{{\rm i} \psi \lambda_{7}} \Gamma {\rm e}^{{\rm i} 
\phi \lambda_{5}} {\rm e}^{{\rm i} \omega \lambda_{2}} \nonumber 
\\
&=&
\left(
\begin{array}{ccc}
  1 & 0 & 0  \\
  0 & c_{\psi} & s_{\psi} \\
  0 & -s_{\psi} & c_{\psi}
\end{array}
\right)
\left(
\begin{array}{ccc}
  1 & 0 & 0  \\
  0 & 1 & 0  \\
  0 & 0 & {\rm e}^{{\rm i} \delta}
\end{array}
\right)
\left(
\begin{array}{ccc}
  c_{\phi} & 0 &  s_{\phi} \\
  0 & 1 & 0  \\
  -s_{\phi} & 0 & c_{\phi}
\end{array}
\right)
\left(
\begin{array}{ccc}
  c_{\omega} & s_{\omega} & 0 \\
  -s_{\omega} & c_{\omega} & 0  \\
  0 & 0 & 1
\end{array}
\right)
\nonumber \\
&=&
\left(
\begin{array}{ccc}
   c_{\phi} c_{\omega} &
   c_{\phi} s_{\omega} &
   s_{\phi}
  \\
   -c_{\psi} s_{\omega}
   -s_{\psi} s_{\phi} c_{\omega} {\rm e}^{{\rm i} \delta} &
   c_{\psi} c_{\omega}
   -s_{\psi} s_{\phi} s_{\omega} {\rm e}^{{\rm i} \delta} &
   s_{\psi} c_{\phi} {\rm e}^{{\rm i} \delta}
  \\
   s_{\psi} s_{\omega}
   -c_{\psi} s_{\phi} c_{\omega} {\rm e}^{{\rm i} \delta} &
   -s_{\psi} c_{\omega}
   -c_{\psi} s_{\phi} s_{\omega} {\rm e}^{{\rm i} \delta} &
   c_{\psi} c_{\phi} {\rm e}^{{\rm i} \delta}
\end{array}
\right),
\label{UPar2}
\end{eqnarray}
where $c_{\psi} = \cos \psi, s_{\phi} = \sin \phi$, etc.

The evolution equation of neutrino with energy $E$
in matter is expressed as
\begin{equation}
 {\rm i} \frac{{\rm d} \nu}{{\rm d} x}
 = H \nu,
 \label{MatEqn}
\end{equation}
where
\begin{equation}
  H
  \equiv
  \frac{1}{2 E}
  \tilde U
  {\rm diag} (\tilde m^2_1, \tilde m^2_2, \tilde m^2_3)
  \tilde U^{\dagger},
 \label{Hdef}
\end{equation}
with a unitary mixing matrix $\tilde U$ and the effective mass squared
$\tilde m^{2}_{i}$'s $(i=1, 2, 3)$.
The matrix $\tilde U$ and the masses
$\tilde m_{i}$'s are determined by\cite{Wolf,MS,BPPW}
\begin{equation}
\tilde U
\left(
\begin{array}{ccc}
  \tilde m^2_1 & & \\
  & \tilde m^2_2 & \\
  & & \tilde m^2_3
\end{array}
\right)
\tilde U^{\dagger}
=
U
\left(
\begin{array}{ccc}
  0 & & \\
  & \delta m^2_{21} & \\
  & & \delta m^2_{31}
\end{array}
\right)
U^{\dagger}
+
\left(
\begin{array}{ccc}
  a & & \\
  & 0 & \\
  & & 0
\end{array}
\right).
\label{MassMatrixInMatter}
\end{equation}
Here $\delta m^2_{ij} = m^2_i - m^2_j$ and
\begin{equation}
 a \equiv 2 \sqrt{2} G_{\rm F} n_{\rm e} E \nonumber \\
   = 7.56 \times 10^{-5} {\rm eV^{2}} \cdot
       \left( \frac{\rho}{\rm g\,cm^{-3}} \right)
       \left( \frac{E}{\rm GeV} \right),
 \label{aDef}
\end{equation}
with the electron density, $n_{\rm e}$ and the averaged
matter density\cite{KS}, $\rho$.
The solution of eq.(\ref{MatEqn}) is then
\begin{eqnarray}
 \nu (x) &=& S(x) \nu(0)
 \label{nu(x)}\\
 S &\equiv& {\rm T\, e}^{ -{\rm i} \int_0^x {\rm d} s H (s) }
 \label{Sdef}
\end{eqnarray}
(T being the symbol for time ordering), giving the oscillation
probability for $\nu_{\alpha} \rightarrow \nu_{\beta} (\alpha, \beta =
{\rm e}, \mu, \tau)$ at distance $L$ as
\begin{eqnarray}
 P(\nu_{\alpha} \rightarrow \nu_{\beta}; E, L)
&=&
 \left| S_{\beta \alpha} (L) \right|^2.
 \label{alpha2beta}
\end{eqnarray}

Note that $P(\bar\nu_{\alpha} \rightarrow \bar\nu_{\beta})$ is related
to $P(\nu_{\alpha} \rightarrow \nu_{\beta})$ through $a \rightarrow
-a$ and $U \rightarrow U^{\ast} ({\rm i.e.\,} \delta \rightarrow
-\delta)$.
Similarly, we obtain $P(\nu_{\beta} \rightarrow \nu_{\alpha})$ from 
eq.(\ref{alpha2beta}) by replacing $\delta \rightarrow -\delta$,
$P(\bar\nu_{\beta} \rightarrow \bar\nu_{\alpha})$ by $a \rightarrow -a$.


Attributing both solar neutrino deficit and atmospheric neutrino 
anomaly to neutrino oscillation, we can assume
$a, \delta m^2_{21} \ll \delta m^2_{31}$.  The oscillation 
probabilities in this case can be considered by perturbation\cite{AKS}.
With the additional conditions
\begin{equation}
    \frac{aL}{2E}
    =
    1.93 \times 10^{-4} \cdot
    \left( \frac{\rho}{\rm g\,cm^{-3}} \right)
    \left( \frac{L}{\rm km} \right)
    \ll 1
    \label{AKScond1}
\end{equation}
and
\begin{equation}
    \frac{\delta m^{2}_{21} L}{2E}
    =
    2.53
    \frac{(\delta m^{2}_{21} / {\rm eV^{2}})(L / {\rm km})}{E / {\rm 
    GeV}} \ll 1,
    \label{AKScond2}
\end{equation}
the matrix $S$ (\ref{Sdef}) is given by
\begin{equation}
 S(x) \simeq {\rm e}^{ -{\rm i} H_0 x } +
             {\rm e}^{ -{\rm i} H_0 x }
              ( -{\rm i} ) \int_{0}^{x} {\rm d} s H_1 (s),
 \label{S0+S1}
\end{equation}
where 
\begin{eqnarray}
  H_0 &=& \frac{1}{2 E} U \left(
 \begin{array}{ccc}
  0 &   &  \\
    & 0 &  \\
    &   & \delta m^2_{31}
 \end{array}
 \right)
 U^{\dagger}
 \label{H0def} \\
 H_1 (x) &=& {\rm e}^{ {\rm i} H_0 x} H_1 {\rm e}^{ -{\rm i} H_0 x},
 \label{H1(x)def}\\
 H_1
&=&
 \frac{1}{2 E}
 \left\{
  U
  \left(
  \begin{array}{ccc}
   0 &                 &  \\
     & \delta m^2_{21} &  \\
     &                 & 0
  \end{array}
  \right)
  U^{\dagger} +
  \left(
  \begin{array}{ccc}
   a &   &  \\
     & 0 &  \\
     &   & 0
  \end{array}
  \right)
 \right\}.
 \label{H1def}
\end{eqnarray}
Then
the oscillation probabilities are calculated, e.g., as
\begin{eqnarray}
& &
 P(\nu_{\mu} \rightarrow \nu_{\rm e}; E, L)
=
 4 \sin^2 \frac{\delta m^2_{31} L}{4 E}
 c_{\phi}^2 s_{\phi}^2 s_{\psi}^2
 \left\{
  1 + \frac{a}{\delta m^2_{31}} \cdot 2 (1 - 2 s_{\phi}^2)
 \right\}
 \nonumber \\
&+&
 2 \frac{\delta m^2_{31} L}{2 E} \sin \frac{\delta m^2_{31} L}{2 E}
 c_{\phi}^2 s_{\phi} s_{\psi}
 \left\{
  - \frac{a}{\delta m^2_{31}} s_{\phi} s_{\psi} (1 - 2 s_{\phi}^2)
  +
 \frac{\delta m^2_{21}}{\delta m^2_{31}} s_{\omega}
    (-s_{\phi} s_{\psi} s_{\omega} + c_{\delta} c_{\psi} c_{\omega})
 \right\}
 \nonumber \\
&-&
 4 \frac{\delta m^2_{21} L}{2 E} \sin^2 \frac{\delta m^2_{31} L}{4 E}
 s_{\delta} c_{\phi}^2 s_{\phi} c_{\psi} s_{\psi} c_{\omega}
 s_{\omega}.
 \label{eq:AKSmu2e}
\end{eqnarray}

As stated, oscillation probabilities such as $P(\bar\nu_{\mu} 
\rightarrow \bar\nu_{\rm e})$, $P(\nu_{\rm e} \rightarrow \nu_{\mu})$
and $P(\bar\nu_{\rm e} \rightarrow \bar\nu_{\mu})$
are given from the above formula by some appropriate changes of the 
sign of $a$ and/or $\delta$.

The first condition (\ref{AKScond1}) of the approximation leads 
to a constraint for the baseline 
length of long-baseline experiments as
\begin{equation}
    L \ll
    1.72 \times 10^{3} {\rm km}
    \left( \frac{\rho}{3 {\rm g\,cm^{-3}}} \right)
    \label{eq:AKScond1'}
\end{equation}
The second condition (\ref{AKScond2}) gives the energy region where we 
can use the approximation,
\begin{equation}
    E
    \gg
    76.0 {\rm MeV}
    \left( \frac{\delta m^{2}_{21}}{10^{-4} {\rm eV^{2}}} \right)
    \left( \frac{L}{300 {\rm km}} \right).
    \label{eq:AKScond2'}
\end{equation}

As long as these conditions, (\ref{eq:AKScond1'}) and
(\ref{eq:AKScond2'}) are satisfied, the approximation (\ref{S0+S1})
works pretty well\cite{AKS,KS2}.

\subsection{Magnitude of CP violation and matter effect}

The available neutrino as an initial beam is 
$\nu_{\mu}$ and $\bar\nu_{\mu}$ in the current
long baseline experiments\cite{K2K,Ferm}.
The ``CP violation'' gives the 
nonzero difference of the oscillation probabilities between, e.g.,
$P(\nu_{\mu} \rightarrow \nu_{\rm e})$ and $
P(\bar\nu_{\mu} \rightarrow \bar\nu_{\rm e})$\cite{AKS}.
This gives
\begin{eqnarray}
 P(\nu_{\mu} \rightarrow \nu_{\rm e}; L)
-
 P(\bar\nu_{\mu} \rightarrow \bar\nu_{\rm e}; L)
&=&
 16 \frac{a}{\delta m^2_{31}} \sin^2 \frac{\delta m^2_{31} L}{4 E}
 c_{\phi}^2 s_{\phi}^2 s_{\psi}^2 (1 - 2 s_{\phi}^2)
 \nonumber \\
&-&
 4 \frac{a L}{2 E} \sin \frac{\delta m^2_{31} L}{2 E}
 c_{\phi}^2 s_{\phi}^2 s_{\psi}^2 (1 - 2 s_{\phi}^2)
 \nonumber \\
&-&
 8 \frac{\delta m^2_{21} L}{2 E}
 \sin^2 \frac{\delta m^2_{31} L}{4 E}
 s_{\delta} c_{\phi}^2 s_{\phi} c_{\psi} s_{\psi} c_{\omega}
 s_{\omega}.
 \label{eq:CP}
\end{eqnarray}
The difference of these two, however, also includes matter effect, or
the fake CP violation, proportional to $a$.  We must somehow
distinguish these two to conclude the existence of CP violation as
discussed in ref.\cite{AKS}.

On the other hand, a muon ring enables to extract $\nu_{\rm e}$ and
$\bar\nu_{\rm e}$ beam.  It enables direct measurement of
pure CP violation through ``T violation'',
e.g., $P(\nu_{\mu} \rightarrow
\nu_{\rm e}) - P(\nu_{\rm e} \rightarrow \nu_{\mu})$ as
\begin{equation}
 P(\nu_{\mu} \rightarrow \nu_{\rm e}) -
 P(\nu_{\rm e} \rightarrow \nu_{\mu})
=
 - 8 \frac{\delta m^2_{21} L}{2 E}
 \sin^2 \frac{\delta m^2_{31} L}{4 E}
 s_{\delta} c_{\phi}^2 s_{\phi} c_{\psi} s_{\psi} c_{\omega}
 s_{\omega}.
 \label{eq:T}
\end{equation}
Note that this difference gives pure CP violation.

By measuring ``CPT violation'', e.g.
the difference between $P(\nu_{\mu} \rightarrow 
\nu_{\rm e})$ and $P(\bar\nu_{\rm e} \rightarrow \bar\nu_{\mu})$,
we can check the matter effect.
\begin{eqnarray}
 P(\nu_{\mu} \rightarrow \nu_{\rm e}; L)
-
 P(\bar\nu_{\rm e} \rightarrow \bar\nu_{\mu}; L)
&=&
 16 \frac{a}{\delta m^2_{31}} \sin^2 \frac{\delta m^2_{31} L}{4 E}
 c_{\phi}^2 s_{\phi}^2 s_{\psi}^2 (1 - 2 s_{\phi}^2)
 \nonumber \\
&-&
 4 \frac{a L}{2 E} \sin \frac{\delta m^2_{31} L}{2 E}
 c_{\phi}^2 s_{\phi}^2 s_{\psi}^2 (1 - 2 s_{\phi}^2)
 \nonumber \\
 \label{eq:CPT}
\end{eqnarray}
%


\begin{figure}
 \unitlength=1cm
 \begin{picture}(15,18)
  \unitlength=1mm
  \centerline{
   \epsfysize=18cm
   \epsfbox{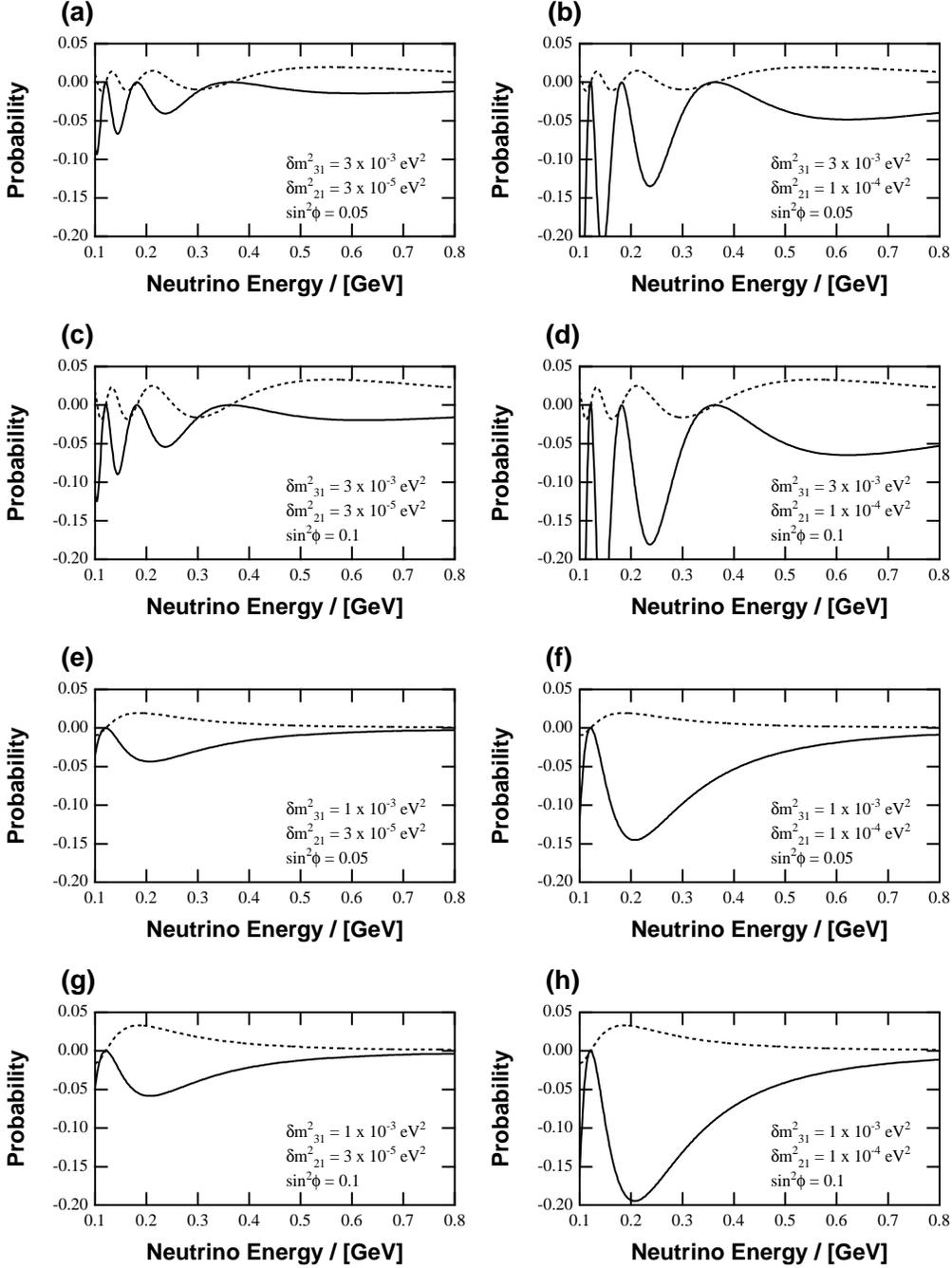}
  }
 \end{picture}
\caption{ Graphs of $P(\nu_{\mu} \rightarrow \nu_{\rm e}) - P(\nu_{\rm
e} \rightarrow \nu_{\mu})$ (solid lines; pure CP-violation effects) and
$P(\nu_{\mu} \rightarrow \nu_{\rm e}) - P(\bar\nu_{\rm e} \rightarrow
\bar\nu_{\mu})$ (dashed lines; matter effects) as functions of neutrino
energy.  Parameters not shown in the graphs are taken as follows.
 $\sin^{2} \omega = 1/2, \sin^{2} \psi = 1/2, \sin \delta
= 1; \rho = {\rm g/cm^{3}}$ and $L = 300 {\rm km}$.} \label{fig1}
\end{figure}


We present in Fig.\ref{fig1} ``T-violation'' part (\ref{eq:T}) and
``CPT-violation'' part (\ref{eq:CPT}) for some parameters allowed by
the present experiments\cite{Fogli}\footnote{Although the Chooz reactor
experiment have almost excluded $\sin^2\phi=0.1$\cite{chooz}, there
remains still small chance to take this value.} with $\sin^{2}
\omega = 1/2$, $\sin^{2} \psi = 1/2$, $\sin \delta = 1$ fixed.  The
matter density is also fixed to the constant value $\rho = 2.5 {\rm
g/cm^{3}}$\cite{KS}.  Other parameters are taken as $\delta m^{2}_{31}
= 3 \times 10^{-3} {\rm eV}^{2}$ and $1 \times 10^{-3} {\rm eV}^{2}$,
$\delta m^{2}_{21} = 1 \times 10^{-4} {\rm eV}^{2}$ and $3 \times
10^{-5} {\rm eV}^{2}$.

``T-violation'' effect is proportional to $\delta m^{2}_{21} / \delta
m^{2}_{31}$ and, for $\phi \ll 1$, also to $\sin \phi$ as seen in
eq.(\ref{eq:T}) and Fig.\ref{fig1}.  Recalling that the energy of neutrino 
beam
is of several hundreds MeV, we see in Fig.\ref{fig1} that the
``T-violation'' effect amounts to at least about 5\%, hopefully
10$\sim$20\%.  This result gives hope to detect the pure leptonic CP
violation directly with the neutrino oscillation experiments.

\section{CP violation in long long baseline experiments}

Here we consider neutrino experiments with baseline $L \sim$
10000km and see that ``T violation'' will be amplified\cite{KP,KS3}.

Since the baseline length $L$ does not
satisfy the condition (\ref{eq:AKScond1'},)
we cannot make use of the previous approximation.

However,
as $a, \delta m^2_{21} \ll \delta m^2_{31}$ is satisfied,
we have approximation formulae of
the mixing matrix in matter $\tilde U$ for {\it neutrino},
\begin{eqnarray}
\tilde U &=&{\rm e}^{{\rm i} \psi \lambda_{7}} \Gamma {\rm e}^{{\rm i} 
\phi \lambda_{5}} {\rm e}^{{\rm i} \tilde\omega \lambda_{2}}
\label{UinMatter}\\
\tan 2\tilde\omega &=& \frac{\delta m^2_{21} s_{2\omega}}
{-a c^2_{\phi}+\delta m^2_{21}c_{2\omega}} 
\nonumber
\end{eqnarray}
and of ``masses'' in matter $\tilde m^2_{i}$ for{\it neutrino}
\begin{eqnarray}
(\tilde m^2_1,\tilde m^2_2,\tilde m^2_3)&=&
(\lambda_-,\lambda_+,\delta m^2_{31}+a c_\phi^2)
\\
\lambda_{\pm}&=&\frac{a c_\phi^2+\delta m^2_{21}}{2}\pm \frac{1}{2}
\sqrt{
(-a c^2_{\psi}+\delta m^2_{21}c_{2\omega})^2+(\delta m^2_{21}s_{2\omega})^2}.
\nonumber
\end{eqnarray}
Thus the ``T violation'' is given by
\begin{eqnarray}
&& P(\nu_e \rightarrow \nu_\mu; L)
-
 P(\nu_{\mu} \rightarrow \nu_{\rm e}; L)
\nonumber\\
&=&4 
 s_{\delta} c_{\phi}^2 s_{\phi} c_{\psi} s_{\psi} c_{\tilde\omega}
 s_{\tilde\omega}\left(
 \sin \frac{\delta \tilde m^2_{21} L}{2 E}+
 \sin \frac{\delta \tilde m^2_{32} L}{2 E}+
 \sin \frac{\delta \tilde m^2_{13} L}{2 E}
\right)
 \label{eq:Tlong}\\
&\sim&
 s_{\delta} c_{\phi}^2 s_{\phi} c_{\psi} s_{\psi} c_{\tilde\omega}
 s_{\tilde\omega}\left(
 \sin \frac{\delta \tilde m^2_{21} L}{2 E}
\right),
\nonumber
\end{eqnarray}
here in the last equation we dropped the terms
$ \sin \frac{\delta \tilde m^2_{32} L}{2 E}+
 \sin \frac{\delta \tilde m^2_{13} L}{2 E}$, since
they oscillate very rapidly and will no contribution to the actual
measurement.

As is seen in eq. (\ref{UinMatter}), due to MSW effect\cite{Wolf,MS}
``T violation'' may be
amplified very much even if the mixing angle $\omega$
is very small and hence we can test whether there is a
CP phase $\delta$\cite{KS3}.

\section{Summary and conclusion}
We considered how large CP/T violation effects can be observed making
use of low-energy neutrino beam, inspired by PRISM\cite{PRISM}.

First we consider the baseline with several hundreds km.
In this case more than 10\%,
hopefully 20\% of the pure CP-violation effects may be observed within
the allowed region of present experiments. To see CP-violation effect
those baseline length and the neutrino energy are most preferable
statistically\cite{KS2}.

Then we consider the baseline with $\sim$ 10000 km.
We see that in this case, due to MSW effect, ``T violation'' will be
amplified and we can test whether the CP phase $\delta$
is large or not.






\end{document}